\documentclass[twocolumn,showpacs,preprintnumbers,amsmath,amssymb,APSl,prd,nofootinbib,superscriptaddress]{revtex4-1}

\usepackage{bm}
\usepackage{mathrsfs}
\usepackage{xcolor,color,graphicx,graphics}
\usepackage[all]{xy}
\usepackage{epsfig,subfigure}
\usepackage{latexsym,amssymb,amsmath,amsfonts} %\usepackage{amstex}
\usepackage[english]{babel} %, spanish, portuguese
\usepackage[OT1]{fontenc}
\usepackage[latin1]{inputenc}
\usepackage{makeidx}
\usepackage{hyperref}
\usepackage{color,graphicx,graphics,wrapfig,epsf}%,psfig
\definecolor{red}{rgb}{1,0,0}

\def\+{^\dagger}

\def\<{\leftarrow}
\def\>{\rightarrow}

\def\({\left(}
\def\){\right)}

\newcommand{\bi}{\begin{itemize}} 				\newcommand{\ei}{\end{itemize}}
\newcommand{\benu}{\begin{enumerate}} 		\newcommand{\enu}{\end{enumerate}}
\newcommand{\bd}{\begin{dinglist}{0}}     \newcommand{\ed}{\end{dinglist}}
\newcommand{\bfig}{\begin{figure}[htbp]}  \newcommand{\efig}{\end{figure}}
        			
\newcommand{\bc}{\begin{center}} 				  \newcommand{\ec}{\end{center}}
\newcommand{\be}{\begin{equation}} 				\newcommand{\ee}{\end{equation}}
\newcommand{\bsub}{\begin{subequations}}  \newcommand{\esub}{\end{subequations}}
\newcommand{\ben}{\begin{eqnarray}} 			\newcommand{\een}{\end{eqnarray}}
\newcommand{\ba}[1]{\begin{array}{#1}} 		\newcommand{\ea}{\end{array}}
\newcommand{\bea}{\begin{equation}\begin{array}{rcl}}
\newcommand{\eea}{\end{array}\end{equation}}

\begin{document}
%\title{Lithium abundance test for modified gravity}
%\title{Lithium abundance is modified gravity dependent}
%\title{Lithium abundance is a modified gravity dependent value}
%\title{Lithium abundance is a modified gravity dependent quantity}
\title{Lithium abundance is a gravitational model dependent quantity}

\author{Aneta Wojnar}
\email{aneta.magdalena.wojnar@ut.ee}
\affiliation{Laboratory of Theoretical Physics, Institute of Physics, University of Tartu,
W. Ostwaldi 1, 50411 Tartu, Estonia
}

\begin{abstract}
The dependence of lithium abundance on modified gravity in low-mass stellar objects is demonstrated. This may introduce an additional uncertainty 
to age determination techniques of young stars and globular clusters if they rely on the light element depletion method.
\end{abstract}

\maketitle
\section{Introduction}
Many alternatives to General Relativity (GR) have been proposed in order to shed light on the dark energy and dark matter 
 problems \cite{Copeland:2006wr,Nojiri:2006ri,nojiri2,nojiri3,Capozziello:2007ec,Carroll:2004de}, the existence of space-time singularities
 \cite{Senovilla:2014gza}, and the unification with the high energy physics \cite{ParTom,BirDav}, between others. Issues related to astrophysical objects also contribute to the above list of current shortcomings of GR, among them for example the
observations of neutron stars with two solar masses \cite{lina,as,craw}, of a compact object with mass $2.6M_\odot$ \cite{NSBH} 
sneaking out the mass bounds given by theoretical models for the heaviest neutron stars and the lightest black holes, and very recently, of a binary black hole merger 
with a total mass of $150M_\odot$ \cite{abotHBH,sak3}.

Pre-main sequence low-mass stars ($M\lesssim0.5M_\odot$) turn out to be interesting objects to study in the context of modified gravity - it was
shown that minimum main sequence mass (MMSM) \cite{sak1,sak2,cris,gonzalo}, cooling process of brown dwarfs \cite{benito}, as well as an upper mass' limit of fully convective stars on the
Main Sequence and Hayashi tracks \cite{aneta4} can be used to constrain theories of gravity. Moreover, during the Hayashi contraction phase 
those young stars
fuse lithium $^7\text{Li}$ which depletes before they reach the Main Sequence - that is,
the temperature required for lithium burning is lower than the one needed for hydrogen fusion. Therefore, the lithium line is not present in spectroscopically observed red 
dwarfs with masses $M\lesssim0.5M_\odot$ in contrast to brown dwarfs whose core temperatures do not reach
 $\sim 2.5\times 10^6~\text{K}$ which is required for lithium burning. This fact, called lithium test \cite{rebolo, nelson}, although not ideal, is used in order to 
 distinguish brown dwarfs from Main Sequence stars in the case when one deals with very low-mass and cool stellar (and substellar) objects occupying overlapping regions of effective temperature and luminosity \cite{bastri}.

The lithium abundance at the photosphere in the pre-main sequence stars is an age-dependent quantity \cite{bastri2,chab,bild,usho}. It allows to determine 
clusters' age in the age range of $20-200\,\text{Myr}$, being one of the most reliable methods for young globular clusters' age determination. It also means that
the lithium depletion boundary method is usually applied to stars' groups of similar age; however, it may provide limits on the ages of LMS
individuals. More importantly, the procedure is employed to calibrate other techniques used for age estimation since the method is built on solid
physical ground, with very few assumptions. Furthermore, the theoretical ages obtained from the lithium depletion method depend weakly 
on stellar compositions, 
which is why they do not provide observational uncertainties \cite{soder} related to, for instance, star's metallicity. Together with keeping the effective temperature as a free parameter the technique allows to avoid further uncertainties related to atmosphere and convection models \cite{bild,usho}. 

Another prominent feature of the pre-main sequence low-mass stars, which we are going to use in this paper, is their theoretical description. Being fully convective, they can be modelled as a well-mixed polytrope with $n=3/2$ even during the last stages of contraction, when electron 
degeneracy starts being important \cite{Burrows:1992fg}. Due to the simplified relations, low-mass stars can be modelled by the non-relativistic hydrostatic equilibrium equation, which turn out to be altered by modified gravity (see \cite{reva} and references therein). That 
fact does not only provide tests for modified gravity as already mentioned, but, what we would like to demonstrate in the following discussion, introduces a new uncertainty to the ages of young stars and globular clusters obtained from the lithium depletion boundary method.

Having this in mind, we are going to demonstrate in this work that the lithium abundance in low-mass stars turn out to be dependent on a gravitational model. In the next section \ref{sec2} we will briefly recall the main features of Palatini $f(\mathcal{R})$ gravity for which we will present the mentioned dependence - however, a similar dependence will also appear in any other model of gravity which alters the stellar description in Newtonian limit. The section \ref{sec3} will provide us the main steps of the derivation of light elements abundance in Palatini gravity and we will examine central temperatures, ages and luminosities of a young low-mass star with respect to the GR model. In the last section \ref{sec4} we will conclude our findings.

Let us add that we use the $(-+++)$ metric signature convention while $\kappa=-\frac{8\pi G}{c^4}$ \cite{weinberg}.

\section{Palatini $f(\mathcal{R})$ gravity}\label{sec2}

Palatini $f(\mathcal{R})$ gravity is one of the simplest generalization of GR - instead of considering the linear Lagrangian of the Ricci scalar, one deals with its general functional, such as:
\begin{equation}
S=S_{\text{g}}+S_{\text{m}}=\frac{1}{2\kappa}\int \sqrt{-g}f(\mathcal{R}) d^4 x+S_{\text{m}}[g_{\mu\nu},\psi_m],\label{action}
\end{equation}
where $\mathcal{R}=\mathcal{R}^{\mu\nu}g_{\mu\nu}$ is the Ricci scalar. It is however constructed with the metric $g_{\mu\nu}$ and Ricci tensor $\mathcal{R}_{\mu\nu}$ built of the independent 
connection $\hat\Gamma$ since the common assumption on $g$-metricity of $\hat\Gamma$ is discarded. 
The field equations are provided by the variation of (\ref{action}) with respect to the metric $g_{\mu\nu}$ 
\begin{equation}
f'(\mathcal{R})\mathcal{R}_{\mu\nu}-\frac{1}{2}f(\mathcal{R})g_{\mu\nu}=\kappa T_{\mu\nu},\label{structural}
\end{equation}
where $T_{\mu\nu}$ is the energy 
momentum tensor of the matter field, $T_{\mu\nu}=-\frac{2}{\sqrt{-g}}\frac{\delta S_m}{\delta g_{\mu\nu}}$.
In the further part of this paper we will assume a perfect fluid form. The prime in (\ref{structural}) denotes the derivative with respect to the 
function's argument, that is, $f'(\mathcal{R})=\frac{df(\mathcal{R})}{d\mathcal{R}}$. On the other hand, the
variation with respect to the independent connection $\hat\Gamma$ gives
\begin{equation}
\hat{\nabla}_\beta(\sqrt{-g}f'(\mathcal{R})g^{\mu\nu})=0,\label{con}
\end{equation}
which demonstrates that $\hat{\nabla}_\beta$ is the covariant derivative calculated with respect to $\hat\Gamma$. In other words, it is the Levi-Civita connection of the conformal metric $h_{\mu\nu}$
\begin{equation}\label{met}
h_{\mu\nu}=f'(\mathcal{R})g_{\mu\nu}.
\end{equation}
The trace equation, obtained by contracting (\ref{structural}) with the metric $g_{\mu\nu}$, is
\begin{equation}
f'(\mathcal{R})\mathcal{R}-2 f(\mathcal{R})=\kappa T,\label{struc}
\end{equation}
where $T$ is the trace of the energy-momentum tensor, and it allows to obtain the relation $\mathcal{R}=\mathcal{R}(T)$ for some chosen functional $f(\mathcal R)$. 

An important feature of the Palatini gravity, which can be derived easily from (\ref{struc}), is that in the vacuum it provides Einstein vacuum solution with the cosmological constant, independently of the form of $f(\mathcal{R})$. Furthermore, in the case of analytic $f(\mathcal{R})$ one deals with the same center-of-mass orbits as in GR \cite{junior}. Therein, it was also demonstrated that the modifications of energy and momentum present in Euler equation are not sensitive to the experiments performed so far for the solar system orbits. This may change when atomic level experiments will be 
available, though \cite{sch,ol1,ol2}.

We may rewrite the field equations (\ref{structural}) as dynamical equations for the conformal 
 metric $ h_{\mu\nu}$ \cite{DeFelice:2010aj,BSS,SSB} and the undynamic scalar 
 field defined as $\Phi=f'(\mathcal{R})$:
 \begin{subequations}
	\begin{align}
	\label{EOM_P1}
	 \bar R_{\mu\nu} - \frac{1}{2} h_{\mu\nu} \bar R  &  =\kappa \bar T_{\mu\nu}-{1\over 2} h_{\mu\nu} \bar U(\Phi)
	%\label{EOM_metric}
%	& \mathcal{\nabla}_\lambda(\sqrt{-g}\Phi g^{\mu\nu})=0
	\end{align}
	\begin{align}
	\label{EOM_scalar_field_P1}
	  \Phi\bar R &  -  (\Phi^2\,\bar U(\Phi))^\prime =0%\\
	% &+ 2\kappa^2 \beta_{,C} T = 0 \,.
	\end{align}
\end{subequations}
where $\bar U(\Phi)=\frac{\mathcal{R}\Phi-f(\mathcal{R})}{\Phi^2}$ and the energy momentum 
tensor in the Einstein frame is given by $\bar T_{\mu\nu}=\Phi^{-1}T_{\mu\nu}$. Let us notice that prime here denotes the derivative with respect to $\Phi$. Such a representation can significantly simplify given physical problems \cite{aneta,o,o1,o2}.

\subsection{Non-relativistic stars in Palatini gravity}\label{substars}

In what follows, we will consider the quadratic (Starobinski) model
\begin{equation}
 f(\mathcal{R})=\mathcal{R}+\beta\mathcal{R}^2,
\end{equation}
where $\beta$ is the model parameter with the dimension $[\rm m^2]$,
for which it was shown that the non-relativistic Palatini stars can be described by the equations \cite{aneta2,gonzalo}
\begin{align}
 \frac{dp}{d\tilde r}&=-\frac{G m(\tilde r)\rho(\tilde r)}{\Phi(\tilde r)\tilde r^2} \ ,\\
 m&=\int_0^{\tilde r}4\pi x^2\rho(x)dx \ ,
\end{align}
where $\tilde r^2=\Phi(\tilde r) r^2$ and $\Phi(\tilde r)\equiv f'(\mathcal{R})=1+2\kappa c^2\beta \rho(\tilde r)$. The transformation to the 
Jordan frame and the Taylor expansion around $\beta=0$ will provide the modified hydrostatic equilibrium equation as
\begin{equation}\label{pres}
 p'=-g\rho(1+\kappa c^2 \beta [r\rho'-3\rho]) \ ,
\end{equation}
where $g=\text{const}$ is the surface gravity assumed to be a constant 
\begin{equation}\label{surf}
 g\equiv\frac{G m(r)}{r^2}\sim\frac{GM}{R^2},
\end{equation}
where $M=m(R)$ and $R$ is a star's radius. On the other hand, the transformation of the mass function $ m(\tilde r)$ to $m(r)$ is dependent on the energy density which will drop however to zero on the
 non-relativistic star's surface. Because of that, the derivation of the mass function has a simple form $ m'(r)=4\pi r^2\rho(r)$ in the Jordan frame, thus we may write
 \begin{equation}
  m''=8\pi r\rho+4\pi r^2 \rho'.
\end{equation}
Using this and (\ref{surf}) in (\ref{pres}), one writes
\begin{equation}\label{hyd}
 p'=-g\rho\left( 1+8\beta\frac{g}{c^2 r} \right).
\end{equation}

As discussed already in the introductory section, our concern are low-mass stars whose equation of state can be modelled by a simple polytropic relation
\begin{equation}\label{pol}
  p=K\rho^\gamma,
\end{equation}
which together with the hydrostatic equilibrium provides the Palatini Lane-Emden equation \cite{aneta2}:
\begin{equation}\label{LE}
 \frac{1}{\xi}\frac{d^2}{d\xi^2}\left[\sqrt{\Phi}\xi\left(\theta-\frac{2\kappa^2 c^2\rho_c\alpha}{n+1}\theta^{n+1}\right)\right]=
 -\frac{(\Phi+\frac{1}{2}\xi\frac{d\Phi}{d\xi})^2}{\sqrt{\Phi}}\theta^n,
\end{equation}
where $\Phi=1+2\alpha \theta^n$ with $\alpha$ defined as $\alpha=\kappa c^2\beta\rho_c$, while the dimensionless variables $\theta$ and $\xi$ are given by
\begin{align}
 r&=r_c\xi,\;\;\;\rho=\rho_c\theta^n,\;\;\;p=p_c\theta^{n+1},\label{def1}\\
 r^2_c&=\frac{(n+1)p_c}{4\pi G\rho^2_c}.\label{def2}
\end{align}
The function $\theta(\xi)$ is the solution of the (modified) Lane-Emden equation (\ref{LE}) with respect to the radial coordinate 
$\xi=r\rho_c\sqrt{8\pi G/(2p_c)}$ which 
crosses zero at $\xi_R$. 
Here, $p_c$ and $\rho_c$ denote the central pressure and density, respectively, while $n=\frac{1}{\gamma-1}$ is the polytropic index of (\ref{pol}). More detailed discussion about Palatini Lane-Emden equation, its solutions, and features can be found in \cite{aneta2,artur,aneta4,aneta3}.

The solution of the modified Lane-Emden equation (\ref{LE}) provide
the star's mass, radius, central density, and temperature via the well-known expressions (see e.g \cite{weinberg})
\begin{align}
 M&=4\pi r_c^3\rho_c\omega_n,\\
 R&=\gamma_n\left(\frac{K}{G}\right)^\frac{n}{3-n}M^\frac{n-1}{n-3} \label{radiuss},\\
 \rho_c&=\delta_n\left(\frac{3M}{4\pi R^3}\right) \label{rho0s} ,\\
 T&=\frac{K\mu}{k_B}\rho_c^\frac{1}{n}\theta_n \label{temps},
\end{align}
where $k_B$ denotes the Boltzmann constant, $\mu$ the mean molecular weight while $K$ contains information about gas mixture and degeneracy of the stellar material. The constants (\ref{omega}) and (\ref{delta}) 
depend on the central energy density via $\Phi$ and its derivation with respect to $\xi$, which is
a common feature of Palatini theories of gravity:
\begin{align}
 \omega_n&=-\frac{\xi^2\Phi^\frac{3}{2}}{1+\frac{1}{2}\xi\frac{\Phi_\xi}{\Phi}}\frac{d\theta}{d\xi}\mid_{\xi=\xi_R},\label{omega}\\
  \gamma_n&=(4\pi)^\frac{1}{n-3}(n+1)^\frac{n}{3-n}\omega_n^\frac{n-1}{3-n}\xi_R,\label{gamma}\\
 \delta_n&=-\frac{\xi_R}{3\frac{\Phi^{-\frac{1}{2}}}{1+\frac{1}{2}\xi\frac{\Phi_\xi}{\Phi}}\frac{d\theta}{d\xi}\mid_{\xi=\xi_R}} \ . \label{delta}
\end{align}
In the next section we will use the above relations to calculate the lithium depletion rate (which can be also easily generalized for resonant rates).

\section{Lithium burning in low-mass main sequence stars}\label{sec3}
As already discussed, fully convective low-mass stars are very-well described by the formalism given in the subsection (\ref{substars}).
In such stars, the lithium-to-hydrogen 
ratio $f$ changes due to the effective convection which mixes lithium-poor and lithium-rich regions throughout a star such that the mixing 
timescale is much shorter than the contraction and lithium destruction times (that is, the star is well mixed). Apart from this process, the proton-capture reactions
also contribute to the rate of change of $f$. Thus, the depletion rate in a star with mass $M$ and hydrogen fraction $X$ can be written in the following way
\begin{equation}\label{reac}
 M\frac{\text{d}f}{\text{d}t}=-\frac{Xf}{m_H}\int^M_0\rho\langle\sigma v\rangle dM,
\end{equation}
where the non-resonant reaction rate for the temperature range $T<6\times 10^6\text{K}$ is given by
\begin{equation}
 N_A\langle\sigma v\rangle=Sf_\text{scr} T^{-2/3}_{c6}\text{exp}\left[-aT_{c6}^{-\frac{1}{3}}\right]\;\frac{\text{cm}^3}{\text{s g}},
\end{equation}
where $T_{c6}\equiv T_c/10^6\text{K}$ and $f_\text{scr}$ is the screening correction factor while $S$ and $a$ are dimensionless parameters 
in the fit to the reaction rate. For our range of temperatures, the proton-capture rate parameters 
for the reaction $^7\text{Li}(p,\alpha)\,^4\text{He}$ are $S=7.2\times10^{10}$ and $a=84.72$ \cite{usho,cf,raimann}. 

Since we are dealing with
polytropic stars with the polytropic index $n=3/2$, the temperature is $T=T_c\theta(\xi)$ while density is expressed as 
$\rho=\rho_c\theta^{3/2}(\xi)$. Therefore, the central temperature $T_c$ and central density $\rho_c$ for that model are modified (via $\delta$, $\xi_R$ and $\theta'$) and given by
\begin{align}\label{ctemp}
 T_c=&1.15\times 10^6 \left(\frac{\mu_\text{eff}}{0.6}\right)\left(\frac{M}{0.1M_\odot}\right) \left(\frac{R_\odot}{R}\right)
 \frac{\delta^\frac{2}{3}}{\xi_R^\frac{5}{3}(-\theta'(\xi_R))^\frac{1}{3}}\text{K}\\
 \rho_c=&0.141\left(\frac{M}{0.1M_\odot}\right) \left(\frac{R_\odot}{R}\right)^3\delta\,\frac{\text{g}}{\text{cm}^3}
\end{align}
while the radius, when taking into account an arbitrary degeneracy degree $\eta$ and mean molecular weight $\mu_\text{eff}$, is
\begin{equation}
 \frac{R}{R_\odot}\approx\frac{7.1\times10^{-2}\gamma}{\mu_\text{eff}\mu_e^\frac{2}{3}F^\frac{2}{3}_{1/2}(\eta)}
 \left(\frac{0.1M_\odot}{M}\right)^\frac{1}{3}\label{Rpol},
\end{equation}
where $F_n(\eta)$ is the $n$th order Fermi-Dirac function. Inserting the Lane-Emden temperature, energy density and radius to (\ref{reac}) and 
changing the variables to the spatial ones we will have 
\begin{align}\label{rate}
  \frac{\text{d}}{\text{d}t}\text{ln}f&=-\frac{4\pi X}{\xi_R^3 }\frac{\rho^2_c R^3}{M}\frac{S}{N_A m_H}\left(\frac{u}{a}\right)^2\nonumber\\
  &\times\int_0^{\xi_R}f_\text{scr} \xi^2\theta^\frac{7}{3} \text{exp}(-u\theta^{-1/3})d\xi\,\,\,\,  \frac{1}{\text{s}},
\end{align}
where $u\equiv aT_6^{-1/3}$. Approximately, the burning process is restricted to the central region of the star, thus we may apply the near 
center solution of the modified Lane-Emden equation (\ref{LE}) to the depletion rate (\ref{rate})
\begin{equation}
 \theta(\xi\approx0) \approx 1-\frac{\xi^2}{6}\approx\text{exp}\left(-\frac{\xi^2}{6}\right),
\end{equation}
which after applying the numerical constants yields
\begin{align}
 \frac{\text{d}}{\text{d}t}\text{ln}f&=-6.54\left(\frac{X}{0.7}\right)\left(\frac{0.6}{\mu_\text{eff}}\right)^3\left(\frac{0.1M_\odot}{M}\right)^2\nonumber\\
 &\times Sf_\text{scr} a^7 u^{-\frac{17}{2}}e^{-u}
 \left(1+\frac{7}{u}\right)^{-\frac{3}{2}}\xi_R^2(-\theta'(\xi_R)).
\end{align}
The integration of the above equation requires the knowledge of the dependence of the central temperature parameter $u$ on time. 
In order to find it, let us consider Stefan-Boltzman equation together with the virial theorem after the transformation to the Jordan frame
\begin{equation}
 L=4\pi R^2 T^4_\text{eff}=-\frac{3}{7}\Omega\frac{GM^2}{R^2}\frac{\text{d}R}{\text{d}t},
\end{equation}
where the factor
\begin{equation}
\Omega=\left(\frac{\Phi^\frac{3}{2}}{1+\frac{1}{2}\xi_R\frac{\Phi'}{\Phi}}\right)^{-\frac{4}{3}}
 \end{equation}
 appears due to the frame transformation\cite{artur}. Therefore, it is straightforward to get the radius and luminosity as functions of time during the contraction phase
\begin{align}
 \frac{R}{R_\odot}=&0.85\Omega^\frac{1}{3} \left(\frac{M}{0.1M_\odot}\right)^\frac{2}{3} \left(\frac{3000\text{K}}{T_\text{eff}}\right)^\frac{4}{3}
 \left(\frac{\text{Myr}}{t}\right)^\frac{1}{3}\label{Rt}\\
  \frac{L}{L_\odot}=& 5.25\times10^{-2}\Omega \left(\frac{M}{0.1M_\odot}\right)^\frac{4}{3} \left(\frac{T_\text{eff}}{3000\text{K}}\right)^{\frac{4}{3}}
 \left(\frac{\text{Myr}}{t}\right)^\frac{2}{3}\label{Lcontr},
\end{align}
while the contraction time is given by
\begin{align}\label{tcon}
 t_\text{cont}\equiv-&\frac{R}{\text{d}R/\text{d}t}\approx841.91  \left(\frac{3000\text{K}}{T_\text{eff}}\right)^4  \left(\frac{0.1M_\odot}{M}\right)\\
 \times&
  \left(\frac{0.6}{\mu_\text{eff}}\right)^3 \left(\frac{T_c}{3\times10^6\text{K}}\right)^3 \frac{\xi_R^2(-\theta'(\xi_R))\Omega}{\delta^2}\,\text{Myr}.\nonumber
\end{align}

From the relation (\ref{Rpol}) and (\ref{Rt}) we may also write down the degeneracy parameter as a function of time
\begin{align}\label{deg}
\mu_\text{eff}F^\frac{2}{3}_{1/2}(\eta)\approx8.36\times10^{-2}\frac{\gamma}{\Omega^{1/3}}
   \left(\frac{0.1M_\odot}{M}\right)    
  \left(\frac{T_\text{3eff}^4 t_6}{\mu^2_e}\right)^{1/3}
\end{align}
where $T_\text{3eff}\equiv T_\text{eff}/3000\text{K}$ and $t_6\equiv t/{10^6}$. Then, using (\ref{ctemp}) together with (\ref{Rpol}),(\ref{Rt}), and (\ref{deg}) we find
\begin{align}
 \frac{u}{a}&=1.15\left(\frac{M}{0.1M_\odot}\right)^{2/9}\left(\frac{\mu_eF_{1/2}(\eta)}{t_6 T^4_{3eff}}\right)^{2/9}\nonumber\\
 &\times
 \left(\frac{\xi_R^5\Omega^{2/3}(-\theta'(\xi_R))^{2/3}}{\gamma\delta^{2/3}}\right)^{1/3},
\end{align}
which relates the central temperature $T_c$ with the time during the contraction phase.

Let us consider the case $M\gtrsim0.2M_\odot$, that is, when the degeneracy effects are not important and $\dot{\mu}_\text{eff}$ can be neglected 
when compared to $\dot{R}$. Then, since $u= aT_{c6}^{-1/3}$ and using (\ref{ctemp}) will provide $\text{d}u/\text{d}R=u/(3R)$, such that
\begin{equation}
 \frac{\text{d}}{\text{d}t}\text{ln}f\approx\frac{\text{dln}f}{\text{d}u} \frac{\partial u}{\partial R}\dot{R}=
 \frac{\text{dln}f}{\text{d}u}\frac{u \dot{R}}{3R}
\end{equation}
which allows to write the depletion rate as
\begin{align}
 &\frac{\text{dln}f}{\text{d}u} = 1.15\times10^{13}~T_\text{3eff}^{-4}\left(\frac{X}{0.7}\right)\left(\frac{0.6}{\mu_\text{eff}}\right)^6
\left(\frac{M_\odot}{M}\right)^3\nonumber\\
&\times Sf_\text{scr} a^{16}u^{-\frac{37}{2}}e^{-u}\left(1-\frac{21}{2u}\right)\frac{\xi_R^4(-\theta'(\xi_R))^2\Omega}{\delta^2}.
 \end{align}
Integrating the above equation from $u_0=\infty$ to $u$ and using the properties of the incomplete gamma function gives
\begin{align}\label{sol}
 \mathcal{F}\equiv\text{ln}\frac{f_0}{f}&=1.15\times10^{13}~T_\text{3eff}^{-4}\left(\frac{X}{0.7}\right)\left(\frac{0.6}{\mu_\text{eff}}\right)^6
\left(\frac{M_\odot}{M}\right)^3\nonumber\\
&\times Sf_\text{scr} a^{16}g(u)\frac{\xi_R^4(-\theta'(\xi_R))^2\Omega}{\delta^2},
\end{align}
where $g(u)=u^{-37/2}e^{-u}-29\Gamma(-37/2,u)$ with the function $\Gamma(-37/2,u)$ being an upper incomplete gamma function. 
For a given depletion $\mathcal{F}$, the central temperature $T_c$ is 
obtained from $u(\mathcal{F})$ while the star's age, radius, and luminosity are specified by the equations (\ref{tcon}), (\ref{Rt}), and (\ref{Lcontr}), respectively. 
Each of those, as demonstrated, is altered by $\Omega$, $\gamma$, $\delta$, $\xi_R$, $\theta(\xi_R)$, and $\theta'(\xi_R)$ whose values depend on the applied model of gravity.

Using the similar approach, one may write down the depletion equation (\ref{sol}) for resonant rates (see e.g (\cite{usho})), where $j=2/3$ stands 
for the non-resonant reactions,
\begin{align}\label{sol2}
 \mathcal{F}\equiv\text{ln}\frac{f_0}{f}&=1.15\times10^{13}~T_\text{3eff}^{-4}\left(\frac{X}{0.7}\right)\left(\frac{0.6}{\mu_\text{eff}}\right)^6
\left(\frac{M_\odot}{M}\right)^3\nonumber\\
&\times Sf_\text{scr} a^{18-3j}g(u)\frac{\xi_R^4(-\theta'(\xi_R))^2\Omega}{\delta^2},
\end{align}
with $g(u)=u^{-41/2-3j}e^{-u}-\frac{68-15j}{2}\Gamma(-\frac{41}{2}-3j,u)$.

The equations (\ref{sol}) and (\ref{sol2}) can be solved numerically or be fitted to the observational data; however, we may also find an approximate formula for the central temperature at the 
time of depletion. Thus, we will compare the local nuclear destruction time at the center of the star ($X=0.7$ being the hydrogen mass fraction while $m_p\approx1.67\times10^{-24}$g is the proton mass)
\begin{align}
 t_\text{dest}&=\frac{m_p}{X\rho\langle\sigma v\rangle}=4.92\times10^{-7}\left(\frac{M}{0.1M_\odot}\right)^2\\
 &\times\left(\frac{\mu_\text{eff}}{0.6}\right)^3
 \frac{T_{c6}^{-\frac{7}{3}}}{Sf_\text{scr}}e^\frac{a}{T_{c6}^{1/3}}\frac{\delta}{\xi_R^5\theta'(\xi_R)}\,\text{yr}\nonumber
\end{align}
to the contraction time (\ref{tcon}). The approximation $t_\text{cont}=t_\text{dest}$ works well so long as the star can be described by the polytropic 
equation of state with $n=3/2$ (the degeneracy is not important):
\begin{align}\label{Ltcen}
 \frac{a}{T_{c6}^{1/3}}&=31.78+\text{ln}(Sf_\text{scr})+\text{ln}\left(\frac{\xi_R^7\theta'(\xi_R)^2\Omega}{\delta^3}\right)-6\text{ln}\left(\frac{\mu_\text{eff}}{0.6}\right)\nonumber\\
 &-3\text{ln}\left(\frac{M}{0.1M_\odot}\right)-4\text{ln}\left(\frac{T_\text{eff}}{3000\text{K}}\right) +\frac{16}{3}\text{ln}T_{c6}.
\end{align}
Let us now consider a star with $T_\text{eff}=3500\text{K}$, the mass $M=0.5M_\odot$ with $f_\text{scr}=1$ (evaluated at the center of the star) 
and $\mu_\text{eff}=0.6$. For the GR values ($\alpha=0$) of the polytropic solutions  $^7\text{Li}$ depletes when the central 
temperature is $T_c\approx2.98\times10^6\text{K}$. A few other values of the parameter $\alpha=\kappa c^2 \beta \rho_c$ and their corresponding central temperatures, ages, radii, and luminosities are given in the table (\ref{tab}). 

The obtained results clearly demonstrate that modified gravity (here Palatini quadratic model) significantly changes the ages and
luminosities of lithium depleted pre-main sequence stars with respect to the GR model. Despite the fact that the values from the table (\ref{tab}) are given by the 
approximated expression (\ref{Ltcen}), the deviations from the GR model with $\alpha=0$ in (\ref{sol2}) are also expected to occur.

\section{Conclusions}\label{sec4}
In this paper we have obtained the lithium-to-hydrogen ratio for the stellar model provided by the Palatini $f(\mathcal{R})$ gravity. Our main result is given by the equations (\ref{sol}), (\ref{sol2}), and (\ref{Ltcen}), where the last one's solutions for a few values of the parameter $\alpha$ are presented in the table (\ref{tab}). All those relations contain terms depending on the solution of (modified) Lane-Emden equation, introducing the dependence on gravitational model of interest. Such a dependence will appear in theories of gravity which modify the Newtonian limit of the relativistic hydrostatic equilibrium equation (that is, the Tolman-Oppenheimer-Volkoff equation).

Although it seems to be worrying that the lithium depletion based techniques for the age estimation depend on the applied model of gravity, it also gives room for gravitational theories whose modifications shorten any phase of the stellar evolution, as provided by the considered Palatini quadratic model. The discovery of a $0.2M_\odot$ white dwarf in the binary system KIC 8145411 \cite{wd} which according to the commonly accepted model would have to be older than the Universe \cite{age}, is a clear example of the need of different evolutionary scenarios (for a brief discussion on that topic, see \cite{aneta5}). It was shown \cite{reid} that white dwarfs are also found in young clusters whose progenitor stars' masses depend crucially on the assumed age of the cluster, which is another argument for being aware of the discussed dependence when the lithium based method is used.

Moreover, staying shorter (longer) in any evolutionary phase has a noticeable effect on the total stars' luminosity which contributes to the galaxy brightness \cite{davis} since the galaxy can have more (less) generations of stars with different luminosities than the ones predicted by the GR model.

In addition, the results discussed in this work may also provide a test for gravitational theories: prolonging prominently low-mass stars' lifetimes in comparison to the current widely accepted model would raise doubts on a theory which introduces such effects. Further studies along these lines are currently underway.

\acknowledgments{The work is supported by the  Regional Development Fund CoE program TK133 ``The Dark Side of the Universe."
 }

\bgroup
\def\arraystretch{1.5}
\begin{table}[ht] 
\begin{center}
\begin{tabular}{|c||c|c|c|c|}
\hline
$\alpha$ & $T_c/10^6\text{K}$ & $t\,\text{[Myr]}$  & $R/R_\odot$ & $L/L_\odot$  \\
\hline\hline %done
-0.4 & 3.48 & 3.21 & 1.85 & 25.3$\times10^{-2}$ \\ 
-0.1 & 3.18 & 7.48 & 1.28 & 14.4$\times10^{-2}$ \\ 
-0.001 & 3.129 & 7.76 & 1.19 & $14.1\times10^{-2}$ \\ 
\hline
0 (GR)    & 2.98 & 12.42 & 1.03 & 10.3$\times10^{-2}$ \\ 
\hline
0.001 & 3.128 & 7.78 & 1.19 & 14$\times10^{-2}$ \\ 
0.1  & 3.098  & 7.25 & 1.13 & 14.7$\times10^{-2}$ \\ 
0.4   & 3.093 & 3.57 & 1.06 & 23.6$\times10^{-2}$ \\  
\hline
\end{tabular}
\caption{Numerical values of central temperatures (in $10^6\text{K}$), age (in Myr), radius (in $R_\odot$), and
luminosity (in $L_\odot$)
of fully convective low-mass stars with respect to 
$\alpha=\kappa c^2 \beta \rho_c$ at the time of $^7\text{Li}$ depletion. The star's mass, effective 
temperature, hydrogen mass fraction, and mean molecular weight are $M=0.5M_\odot$, $T_\text{eff}=3500\text{K}$, $X=0.7$, and $\mu_\text{eff}=0.6$, respectively.
}
\label{tab}
\end{center}
\end{table}

\end{document}